*Vers 02 Nov 2016*

**A State Space Approach for Piecewise-Linear Recurrent Neural Networks for Reconstructing Nonlinear Dynamics from Neural Measurements**

*Short title*: Nonlinear State Space Model for Reconstructing Computational Dynamics

Daniel Durstewitz

Dept. of Theoretical Neuroscience, Bernstein Center for Computational Neuroscience Heidelberg-Mannheim, Central Institute of Mental Health, Medical Faculty Mannheim/ Heidelberg University

daniel.durstewitz@zi-mannheim.de

**Abstract**

The computational and cognitive properties of neural systems are often thought to be implemented in terms of their network dynamics. Hence, recovering the system dynamics from experimentally observed neuronal time series, like multiple single-unit recordings or neuroimaging data, is an important step toward understanding its computations. Ideally, one would not only seek a (lower-dimensional) state space representation of the dynamics, but would wish to have access to its (computational) governing equations for in-depth analysis. Recurrent neural networks (RNNs) are a computationally powerful and dynamically universal formal framework which has been extensively studied from both the computational and the dynamical systems perspective. Here we develop a semi-analytical maximum-likelihood estimation scheme for piecewise-linear RNNs (PLRNNs) within the statistical framework of state space models, which accounts for noise in both the underlying latent dynamics and the observation process. The Expectation-Maximization algorithm is used to infer the latent state distribution, through a global Laplace approximation, and the PLRNN parameters iteratively. After validating the procedure on toy examples, and using inference through particle filters for comparison, the approach is applied to multiple single-unit recordings from the rodent anterior cingulate cortex (ACC) obtained during performance of a classical working memory task, delayed alternation. A model with 5 states turned out to be sufficient to capture the essential computational dynamics underlying task performance, including stimulus-selective delay activity. The estimated models were rarely multi-stable, however, but rather were tuned to exhibit slow dynamics in the vicinity of a bifurcation point. In summary, the present work advances a semi-analytical (thus reasonably fast) maximum-likelihood estimation framework for PLRNNs that may enable to recover the computationally relevant dynamics underlying observed neuronal time series, and directly link them to computational properties.

**Author Summary**

Neuronal dynamics mediate between the physiological and anatomical properties of a neural system and the computations it performs, in fact may be seen as the 'computational language' of the brain. It is therefore of great interest to recover from experimentally recorded time series, like multiple single-unit or neuroimaging data, the underlying network dynamics and, ideally, even its governing equations. This is not at all a trivial enterprise, however, since neural systems are very high-dimensional, come with considerable levels of intrinsic (process) noise, are usually only partially observable, and these observations may be further corrupted by noise from measurement and preprocessing steps. The present article embeds piecewise-linear recurrent neural networks (PLRNNs) within a state space approach, a statistical estimation framework that deals with both process and observation noise. PLRNNs are computationally and dynamically powerful model systems. Their statistically principled estimation from multivariate neuronal time series thus may provide access to some essential features of the neuronal dynamics, like attractor states, their governing equations, and their computational implications. The approach is exemplified on multiple single-unit recordings from the rat prefrontal cortex during working memory.

**Introduction**

Neural dynamics mediate between the underlying biophysical and physiological properties of a neural system and its computational and cognitive properties (e.g. [1-4]). Hence, from a computational perspective, we are often interested in recovering the neural network dynamics of a given brain region or neural system from experimental measurements. Yet, experimentally, we commonly have access only to noisy recordings from a relatively small proportion of neurons (compared to the size of the brain area of interest), or to lumped surface signals like local field potentials or the EEG. Inferring from these the computationally relevant underlying dynamics is therefore not trivial, especially since both the neural system itself (e.g., stochastic synaptic release; [5]) as well as the recorded signals (e.g., spike sorting errors; [6]) come with a good deal of noise.

Speaking in statistical terms, 'model-free' techniques which combine state space reconstruction methods (delay embeddings) from nonlinear dynamics with nonlinear basis expansions and kernel techniques have been one approach to the problem [7, 8]. These techniques provide informative lower-dimensional visualizations of population trajectories and approximations to the neural flow field, but they highlight only certain, salient aspects of the dynamics and do not return its governing equations (e.g. [9]) or underlying computations. Alternatively, state space models, a statistical framework particularly popular in engineering and ecology (e.g. [10]), have been adapted to extract lower-dimensional neural trajectory flows from higher-dimensional recordings [11-21]. State space models link a process model of the unobserved (latent) underlying dynamics to the experimentally observed time series via observation equations, and differentiate between process noise and observation noise (e.g. [22]). So far, with few exceptions (e.g. [19, 23]), these models assumed linear latent dynamics, however. Although this may be sufficient to yield smoothed trajectories and reduced state space representations, it implies that the recovered dynamical model by itself is not powerful enough to reproduce a range of important dynamical and computational phenomena in the nervous system, among them multi-stability which has been proposed to underlie neural activity during working memory [24-28].

Here we derive a new state space algorithm based on piecewise-linear (PL) recurrent neural networks (RNN). It has been shown that RNNs with nonlinear activation functions can, in principle, approximate any dynamical system's trajectory or, in fact, dynamical system itself (given some general conditions; [29-31]). Thus, in theory, they are powerful enough to recover whatever dynamical system is underlying the experimentally observed time series. Piecewise linear activation functions, in particular, are by now the most popular choice in deep learning algorithms [32, 33], and considerably simplify some of the derivations within the state space framework (as shown later). They may also be more apt for producing working memory-type activity with longer delays if for some units the transfer function happens to coincide with the bisectrix (cf. [34]), and ease the analysis of fixed points and stability. We then apply this newly derived algorithm to multiple single-unit recordings from the rat prefrontal cortex obtained during a classical delayed alternation working memory task [35].

**Results**

*State space model*

This article considers simple discrete-time piecewise-linear (PL) recurrent neural networks (RNN) of the form

(1)    $\mathbf{z}_t = \mathbf{A}\mathbf{z}_{t-1} + \mathbf{W}\max\{\mathbf{0}, \mathbf{z}_{t-1} - \boldsymbol{\theta}\} + \mathbf{s}_t + \boldsymbol{\varepsilon}_t \quad , \boldsymbol{\varepsilon}_t \sim N(\mathbf{0}, \boldsymbol{\Sigma})$,

where $\mathbf{z}_t = (z_{1t}...z_{Mt})^T$ is the (M×1)-dimensional (latent) neural state vector at time t=1…T, $\mathbf{A} = diag([a_{11}...a_{MM}])$ is an M×M diagonal matrix of auto-regression weights, $\mathbf{W} = (0\ w_{12}...w_{1M}, w_{21}\ 0\ w_{23}...w_{2M}, w_{31}\ w_{32}\ 0\ w_{34}...w_{3M},...)$ is an M×M *off*-diagonal matrix of connection weights, $\boldsymbol{\theta} = (\theta_1...\theta_M)^T$ is a set of (constant) activation thresholds, $\mathbf{s}_t$ is a sequence of (known) external inputs, and $\boldsymbol{\varepsilon}_t$ denotes a Gaussian white noise process with diagonal covariance matrix $\boldsymbol{\Sigma} = diag([\sigma_{11}^2...\sigma_{MM}^2])$. The max-operator is assumed to work element-wise.

Before proceeding further, two things are worth pointing out: First, more complicated PL functions may, in principle, be constructed from (1) by properly connecting simple PL units, combined with an appropriate choice of activation thresholds θ (and acknowledging the activation lags among units). Second, all fixed points (in the absence of external input) of the PLRNN (1) could be obtained by solving the $2^M$ linear equations

(2)    $\mathbf{z}_* = (\mathbf{A} + \mathbf{W}_\Omega - \mathbf{I})^{-1}\mathbf{W}_\Omega\boldsymbol{\theta}$ ,

where Ω is to denote the set of indices of units for which we assume $z_m \leq \theta_m$, and $\mathbf{W}_\Omega$ the respective connectivity matrix in which all columns from **W** corresponding to units $\in \Omega$ are set to 0. Obviously, to make $\mathbf{z}_*$ a true fixed point of (1), the solution to (2) has to be consistent with the defined set Ω, that is $z_{*m} \leq \theta_m$ has to hold for all $m \in \Omega$ and $z_{*m} > \theta_m$ for all $m \notin \Omega$. For networks of moderate size (say M<30) it is thus computationally feasible to explicitly check for all fixed points and their stability.

Here, latent state model (1) is then connected to some N-dimensional observed vector time series **X**={**x**$_t$} via a simple linear-Gaussian model,

(3)    $\mathbf{x}_t = \mathbf{B}\phi(\mathbf{z}_t) + \boldsymbol{\eta}_t \quad , \boldsymbol{\eta}_t \sim N(\mathbf{0}, \boldsymbol{\Gamma})$,

where $\phi(\mathbf{z}_t) := \max\{\mathbf{0}, \mathbf{z}_t - \boldsymbol{\theta}\}$, $\{\boldsymbol{\eta}_t\}$ is the (white Gaussian) observation noise series with diagonal covariance matrix $\boldsymbol{\Gamma} = diag([\gamma_{11}^2...\gamma_{NN}^2])$, and **B** an N×M matrix of regression weights. Thus, the idea is that only the PL-transformed activation $\phi(\mathbf{z}_t)$ reaches the 'observation surface' (as, e.g., with spiking activity when the underlying membrane dynamics itself is not visible). We further assume for the initial state,

(4)    $\mathbf{z}_1 \sim N(\boldsymbol{\mu}_0 + \mathbf{s}_1, \boldsymbol{\Sigma})$,

with, for simplicity, the same covariance matrix as for the process noise in general (reducing the number of to be estimated parameters). In the case of multiple, temporally separated trials, we allow each one to have its own individual initial condition $\boldsymbol{\mu}_k$, $k = 1...K$.

The general goal here is to determine both the model's unknown parameters $\Xi = \{\boldsymbol{\mu}_0, \mathbf{A}, \mathbf{W}, \boldsymbol{\Sigma}, \mathbf{B}, \boldsymbol{\Gamma}\}$ (assuming fixed thresholds θ for now) as well as the unobserved, latent state path $\mathbf{Z} := \{\mathbf{z}_t\}$ (and its second-order moments) from the experimentally observed time series {$\mathbf{x}_t$}. These could be, for instance, properly transformed multivariate spike time series or neuroimaging data. This is accomplished here by the Expectation-Maximization (EM) algorithm which iterates state (E) and parameter (M) estimation steps and is developed in detail for model (1), (3), in the Methods. In the following I will first discuss state and parameter estimation separately for the PLRNN, before describing results from the full EM algorithm in subsequent sections. This will be done along two toy problems, an higher-order nonlinear oscillation (stable limit cycle), and a simple 'working memory' paradigm in which one of two discrete stimuli had to be retained across a temporal interval. Finally, the application of the validated PLRNN EM algorithm will be demonstrated on multiple single-unit recordings obtained from rats on a standard working memory task (delayed alternation; data from [35], kindly provided by Dr. James Hyman, University of Nevada, Las Vegas).

*State estimation*

The latent state distribution, as explained in Methods, is a high-dimensional (piecewise) Gaussian mixture with the number of components growing as $2^{T \times M}$ with sequence length T and number of latent states M. Here a semi-analytical, approximate approach was developed that treats state estimation as a combinatorial problem by first searching for the mode of the full distribution (cf. [36, 37]; in contrast, e.g., to a recursive filtering-smoothing scheme that makes local [linear-Gaussian] approximations, e.g. [11], cf. [22]). This approach amounts to solving a high ($2^{M \times T}$)-dimensional piecewise linear problem (due to the piecewise quadratic, in the states **Z**, log-likelihood eq. 6, 7). Here this was accomplished by alternating between (1) solving the linear set of equations implied by a given set of linear constraints $\boldsymbol{\Omega} := \{(m,t) \,|\, z_{mt} \leq \theta_m\}$ (cf. eq. 7 in Methods) and (2) flipping the sign of the constraints violated by the current solution $\mathbf{z}_*(\boldsymbol{\Omega})$ to the linear equations, thus following a path through the (M×T)-dimensional binary space of linear constraints using Newton-type iterations (similar as in [38], see Methods). Given the mode and state covariance matrix (evaluated at the mode from the negative inverse Hessian), all other expectations needed for the EM algorithm were then derived analytically, with one exception that was approximated (see Methods for full details).

The toy problems introduced above were used to assess the quality of these approximations. For the first toy problem, an order-15 limit cycle was produced with a PLRNN consisting of three recurrently coupled units, inputs to units #1 and #2, and parameter settings as indicated in Fig. 1 and provided Matlab file 'PLRNNoscParam.mat'. The limit cycle was repeated for 50 full cycles (giving 750 data points) and corrupted by process noise (cf. Fig. 1). These noisy states (arranged in a (3 x 750) matrix **Z**) were then transformed into a (3 x 750) output matrix **X**, to which observation noise was added, through a randomly drawn (3 x 3) regression weight matrix **B**. State estimation was started from a random initial condition. True (but noise-corrupted) and estimated states for this particular problem are illustrated in Fig. 1A, indicating a tight fit (although some fraction of the linear constraints were still violated, ≈0.27% in the present example and <2.3% in the working memory example below; see Methods on this issue).

To examine more systematically the quality of the approximate-analytical estimates of the first and second order moments of the joint distribution across states *z* and their piecewise linear transformations ϕ(*z*), samples from p(**Z**|**X**) were simulated using bootstrap particle filtering (see Methods). Although these simulated samples are based only on the filtering (not the smoothing) steps (and (re-)sampling schemes may have issues of their own; e.g. [22]), analytical and sampling estimates were in tight agreement, correlating almost to 1 for this example, as shown in Fig. 2.

Fig. 3A illustrates the setup of the 'two-cue working memory task', chosen for later comparability with the experimental setup. A 5-unit PLRNN was first trained by conventional gradient descent ('real-time recurrent learning' (RTRL), see [39, 40]) to produce a series of six 1's on unit #3 and six 0's on unit #4 five time steps after an input (of 1) occurred on unit #1, and the reverse pattern (six 0's on unit #3 and six 1's on unit #4) five time steps after an input occurred on unit #2. A stable PLRNN with a reasonable solution to this problem was then chosen for further testing the present algorithm (cf. Fig. 3C). (While the RTRL approach was chosen to derive a working memory circuit with reasonably 'realistic' characteristics like a wider distribution of weights, it is noted that a multi-stable network is relatively straightforward to construct explicitly given the analytical accessibility of fixed points [see Methods]; for instance, choosing $\boldsymbol{\theta} = (0.5\ 0.5\ 0.5\ 0.5\ 2)$, $\mathbf{A} = (0.9\ 0.9\ 0.9\ 0.9\ 0.5)$, and $\mathbf{W} = (0\ \omega\ -\omega\ -\omega\ -\omega, \omega\ 0\ -\omega\ -\omega\ -\omega, -\omega\ -\omega\ 0\ \omega\ -\omega, -\omega\ -\omega\ \omega\ 0\ -\omega, 1\ 1\ 1\ 1\ 0)$ with $\omega = 0.2$, yields a tri-stable system.) Like for the limit cycle problem before, the number of observations was taken to be equal to the number of latent states, and process and observation noise were added (see Fig. 4 and Matlab file 'PLRNNwmParam.mat' for specification of parameters). The system was simulated for 20 repetitions of each trial type (i.e., cue-1 or cue-2 presentations) with different noise realizations and each 'trial' started from its own initial condition $\boldsymbol{\mu}_k$ (see Methods), resulting in a total series length of T=20×2×20=800 (although, importantly, in this case the time series consisted of distinct, temporally segregated trials, instead of one continuous series, and was treated as such an ensemble of series by the algorithm). As before, state estimation started from random initial conditions and was provided with the correct parameters, as well as with the observation matrix **X**. While Fig. 3B illustrates the correlation between true (i.e., simulated) and estimated states across all trials and units, Fig. 3C shows true and estimated states for a representative cue-1 (left) and cue-2 (right) trial, respectively. Again, our procedure for obtaining the maximum a-posteriori (MAP) estimate of the state distribution appears to work quite well (in general, only locally optimal solutions can be guaranteed, however, and the algorithm may have to be repeated with different state initializations; see Methods).

*Parameter estimation*

Given the true states, how well would the algorithm retrieve the parameters of the PLRNN? To assess this, the actual model states (which generated the observations **X**) from simulation runs of the oscillation and the working memory task described above were provided as initialization for the E-step. Based on these, the algorithm first estimated the state covariances for *z* and ϕ(*z*) (see above), and then the parameters in a second step (i.e., the M-step). Note that the parameters can all be computed analytically given the state distribution (see Methods), and, provided the state covariance matrices (summed across time) as required in eqn. 17a,d,f are non-singular, have a unique solution. Hence, in this case, any misalignment with the true model parameters can only come from one of two sources: i) estimation was based on one finite-length noisy realization of the PLRNN process, ii) all *second order moments* of the state distribution were still *estimated* based on the true state vectors. However, as can be appreciated from Fig. 1B (oscillation) and Fig. 4 (working memory), for the two example scenarios studied here, all parameter estimates still agreed tightly with those describing the true underlying model.

In the more general case where *both* the states and the parameters are unknown and only the observations are given, note that the model as stated in eqns. 1 & 3 is over-specified as, for instance, at the level of the observations, additional variance placed into $\Sigma$ can be compensated for by adjusting $\Gamma$ accordingly (cf. [41, 42]). In the following we therefore always arbitrarily fixed $\Sigma = \mathbf{I} \cdot 10^{-2}$, as common in many latent variable models (like factor analysis), including state space models (e.g. [23, 43]).

*Joint estimation of states and parameters by EM*

The observations above confirm that our algorithm finds satisfactory approximations to the underlying state path and state covariances when started with the right parameters, and - vice versa - identifies the correct parameters when provided with the true states. Indeed, the M-step, since it is exact, can only increase the expected log-likelihood eq. 5 with the present state expectancies fixed. However, due to the system's piecewise-defined discrete nature, modifying the parameters may lead to a new set of constraint violations, that is may throw the system into a completely different linear subspace which may imply a decrease in the likelihood in the E-step. It is thus not guaranteed that a straightforward EM algorithm converges (cf. [44, 45]), or that the likelihood would even monotonically increase with each EM iteration.

To examine this issue, full EM estimation of the WM model (as specified in Fig. 4, using *N*=20 outputs in this case) was run 240 times, starting from different random, uniformly distributed initializations for the parameters. Fig. 5B gives, for the maximum likelihood solution across all 240 runs (Fig. 5A), the correlations between true and estimated states for all five state variables of the WM model. Note that estimated and true model states may not be in the same order, as any permutation of the latent state indices together with the respective columns of observation matrix **B** will be equally consistent with the data **X** (see also [23]). For the WM model examined here, however, partial order information is implicitly provided to the EM algorithm through the definition of unit-specific inputs $s_{it}$. For the present example, true and estimated states were nicely linearly correlated for all 5 latent variables (Fig. 5B), but some of the regression slopes significantly differed from 1, indicating a degree of freedom in the scaling of the states. More generally, there may not even be a clear linear relationship with a *single* latent state, although, if the estimation was successful, a linear transformation $\mathbf{V}\hat{\mathbf{Z}}$ may usually map the estimated onto the true states. This is because, if the observation eq. were strictly linear, any linear transformation of the latent states by some matrix **V** could essentially be reversed at the level of the outputs by back-multiplying **B** with $\mathbf{V}^{-1}$ (cf. [23]; note that here the *piecewise linearity* through $\phi(\mathbf{z})$ in eq. 1, 3, ignored in the argument above, complicates matters, however). This implies that the model is (at most) identifiable only up to this linear transformation, which might not be a serious issue, however, if one is interested primarily in the latent dynamics (rather than in the exact parameters).

Fig. 6 illustrates the distribution of initial and final parameter estimates around their true values across all 240 runs (before and after reordering the estimated latent states based on the rotation that would be required for achieving the optimal mapping onto the true states, as determined through Procrustes analysis). Fig. 6 reveals that a) the EM algorithm does clearly improve the estimates and b) these final estimates seemed to be largely unbiased (deviations centered around 0).

*Application to experimental recordings*

I next was interested in what kind of structure the present PLRNN approach would retrieve from experimental multiple (*N*=19) single-unit recordings obtained while rats were performing a simple and well-examined working memory task, namely spatial delayed alternation [35] (see Methods). (Note that in the present context this analysis is mainly meant as an exemplification of the current model approach, not as a detailed examination of the working memory issue itself.) The delay was always initiated by a nose poke of the animal into a port located on the side opposite from the response levers, and had a minimum length of 10 s. Spike trains were first transformed into kernel density estimates by convolution with a Gaussian kernel (see Methods), as done previously (e.g. [8, 46, 47]), and binned with 500 ms resolution. This also renders the observed data more suitable to the Gaussian noise assumptions of the present observation model, eq. 3. Models with 5 and 8 latent states were estimated, but only results from the former will be reported here (as 8 states did not appear to yield any additional insight). Periods of cue presentation were indicated to the model by setting external inputs $s_{it} = 1$ to units *i*=1 (left lever) or *i*=2 (right lever) for three 500 ms time bins surrounding the event (and $s_{it} = 0$ otherwise). The EM algorithm was started from 36 different initializations of the parameters (including thresholds θ), and the 5 highest likelihood solutions were considered further.

Fig. 7A gives the model log-likelihoods across EM iterations for these 5 highest-likelihood solutions. Interestingly, there were single neurons whose responses were predicted very well by the estimated model despite large trial-to-trial fluctuations (Fig. 7B, top row), while there were others with similar trial-to-trial fluctuations for which the model only captured the general trend (Fig. 7B, bottom row). This could potentially suggest that trial-to-trial fluctuations in single neurons could be for very different reasons: In those cases where strongly varying single unit responses are nevertheless highly predictable, at least a considerable proportion of their trial-to-trial fluctuations must have been captured by the deterministic part of the model's latent state dynamics, hence may be due to different (trial-unique) initializations of the states (recall that the states are not free to vary in accounting for the observations, but are tightly constrained by the model's temporal consistency requirements). In contrast, when only the average trend is captured, the neuron's trial-to-trial fluctuations likely represent true intrinsic (or measurement) noise sources that the model's deterministic part cannot account for. This observation highlights that (nonlinear) state space models could potentially also provide new insights into other long-standing questions in neurophysiology.

Fig. 8 shows the five trial-averaged latent states for both left- and right-lever trials for one of the highest likelihood solutions. Not surprisingly, the first two state variables (receiving external input) exhibit a strong cue response for the left vs. right lever, respectively. The third latent variable appears to reflect the (end-of-trial) motor response, while the fourth and fifth state variable clearly distinguish between the left and right lever options throughout the delay period of the task, in this sense carrying a memory of the cue (previous response) within the delay. Hence, for this particular data set, the extracted latent states appear to summarize quite well the most salient computational features of this simple working memory task.

Further insight might be gained by examining the system's fixed points and their eigenvalue spectrum. For this purpose, the EM algorithm was started from 200 different initial conditions (that is, initial parameter estimates and threshold settings θ) with maximum absolute eigenvalues (of the corresponding fixed points) drawn from a relatively uniform distribution within the interval [0 3].

Although the estimation process rarely returned truly multi-stable solutions (just 2% of all cases), there was a clear trend for the final maximum eigenvalues to aggregate around 1 (Fig. 9), that is to produce models with very slow dynamics. Indeed, effectively slow dynamics is all that is needed to bridge the delays (see also [1]), while true multi-stability may perhaps even be the physiologically less likely scenario (e.g. [48, 49]). (Reducing the bin width from 500 ms to 100 ms appeared to produce solutions with eigenvalues even closer to 1 while retaining stimulus selectivity across the delay, but this observation was not followed up more systematically here).

**Discussion**

*Reconstructing neuronal dynamics parametrically and non-parametrically*

In the present work, a semi-analytical, maximum-likelihood (ML) approach for estimating piecewise-linear recurrent neural networks (PLRNN) from brain recordings was developed. The idea is that such models would provide 1) a representation of neural trajectories and computationally relevant dynamical features underlying high-dimensional experimental time series in a much lower-dimensional latent variable space (cf. [16, 19]), and 2) more direct access to the neural system's computational properties. Specifically, once estimated to reproduce the data (in the ML sense), such models may allow for more detailed analysis and in depth insight into the system's computational dynamics, e.g. through an analysis of fixed points and their linear stability (e.g. [24, 26, 28, 50-56]), which is not directly accessible from the experimental time series.

Model-free (non-parametric) techniques, usually based on Takens' delay embedding theorem [57] and extensions thereof [58, 59], have also frequently been applied to gain insight into neuronal dynamics and its essential features, like attracting states associated with different task phases from in-vivo multiple single-unit recordings [7, 8] or unstable periodic orbits extracted from relatively low-noise slice recordings [60]. In neuroscience, however, one commonly deals with high-dimensional observations, as provided by current multiple single-unit or neuroimaging techniques (which still usually constitute just a minor subset of all the system's dynamical variables). In addition, there is a large variety of both process and measurement noise sources. The former include potential thermal noise sources and the probabilistic behavior of single ion channel gating [61], probabilistic synaptic release [5], fluctuations in neuromodulatory background and hormone levels, and a large variety of uncontrollable external noise sources via the sensory surfaces, including somatosensory and visceral feedback from within the body. Measurement noise may come from direct physical sources like, for instance, instabilities and movement in the tissue surrounding the recording electrodes, noise properties of the recording devices themselves, the mere fact that only a fraction of all system variables is experimentally accessed ('sampling noise'), or may result from preprocessing steps like spike sorting (e.g. [62, 63]). This is therefore a quite different scenario from the comparatively low-dimensional and low-noise situations in, e.g., laser physics [64], and delay-embedding-based approaches to the reconstruction of neural dynamics may have to be augmented by machine learning techniques to retrieve at least some of its most salient features [7, 8].

Of course, model-based approaches like the one developed here are also plagued by the high dimensionality and high noise levels inherent in neural data, but perhaps to a lesser extent than approaches like delay embeddings that aim to directly construct the state space from the observations (see also [65]). This is because models as pursued in the statistical state space framework explicitly incorporate both process and measurement noise into the system's description. Also, as long as the latent variable space itself is relatively small and related to the observations by simple linear equations, as here, the high dimensionality of the observations themselves does not constitute a too serious issue for estimation. More importantly, however, it is of clear advantage to have access to the governing equations themselves, as only this allows for an in depth analysis of the system's dynamics and its relation to neural computation (e.g. [2, 24, 26, 53, 54, 56, 66]). For instance, recurrent network models have been trained in the past to perform behavioral tasks or reproduce behavioral data to infer the dynamical mechanisms potentially underlying working memory [67] or context-dependent decision making [54], although commonly, as in the cited cases, not within a statistical framework (or not even by direct estimation from experimental data). There are also approaches which are somewhat in between, attempting to account for the observations directly, without reference to an underlying latent variable model, through differential equations expressed in terms of nonlinear basis expansions in the observations, estimated through strongly regularized (penalized) regression methods ([7], see also [9]). It remains to be investigated how well such methods, which go without a noise model and face high data dimensionality directly, transfer to neuroscience problems.

*Estimation of neural state space models*

State space models are a popular statistical tool in many fields of science (e.g. [10, 68]), although their applications in neuroscience are of more recent origin [11, 12, 14, 17-19, 21]. The Dynamic Causal Modeling (DCM) framework advanced in the human fMRI literature to infer the functional connectivity of brain networks and their dependence on task conditions [68, 69] may be seen as a state space approach, although these models usually do not contain process noise (except for the recently proposed 'stochastic DCM' [69]) and are commonly estimated through Bayesian inference, which imposes more constraints (via computational burden) on the complexity of the models that can reasonably be dealt with in this framework. In neurophysiology, Smith & Brown [11] were among the first to suggest a state space model for multivariate spike count data by coupling a linear-Gaussian transition model with Poisson observations, with state estimation achieved by making locally Gaussian approximations to eq. 18. Similar models have variously been used subsequently to infer local circuit coding properties [14] or, e.g., biophysical parameters of neurons or synaptic inputs from postsynaptic voltage recordings [70, 12]. Yu et al. [21] proposed Gaussian Process Factor Analysis (GPFA) for retrieving lower-dimensional, smooth latent neural trajectories from multiple spike train recordings. In GPFA, the correlation structure among the latent variables is specified (parameterized) explicitly rather than being given through a transition model. Buesing et al. [16] discuss regularized forms of neural state space models to enforce their stability, while Macke et al. [18] review different estimation methods for such models like the Laplace approximation or variational inference methods.

By far most of the models discussed above are linear in their latent dynamics, however (although observations may be non-Gaussian). Although this may often be sufficient to uncover important properties of underlying latent processes or structures, like connectivity or synaptic/neuronal parameters, or to obtain lower-dimensional representations of the observed process, it is not suitable for retrieving the system dynamics or computations, as linear systems are strongly limited in the repertoire of dynamics (and computations) they can produce (e.g. [50, 71]). There are a few exceptions, however, the current work builds on: Yu et al. [19] suggested a RNN with sigmoid-type activation function (using the error function), coupled to Poisson spike count outputs, and used it to reconstruct the latent neural dynamics underlying motor preparation and planning in non-human primates. In their work, they combined the Gaussian approximation suggested by Smith & Brown [11] with the Extended Kalman Filter (EKF) for estimation within the EM framework. These various approximations in conjunction with the iterative EKF estimation scheme may be quite prone to numerical instabilities and accumulating errors, however (cf. [22]). Earlier work by Roweis & Ghahramani [23] used radial basis function (RBF) networks as a partly analytically tractable approach. Nonlinear extensions to DCM, incorporating quadratic terms, have been proposed as well recently [72]. State and parameter estimation has also been attempted in (noisy) nonlinear biophysical models [73,74], but these approaches are usually computationally expensive, especially when based on numerical sampling [74], while at the same time pursuing objectives somewhat different from those targeted here (i.e., less focused on computational properties).

Nevertheless, nonlinear neural state space models remain an under-researched topic in theoretical neuroscience. In the present work, PLRNNs were therefore chosen as a mathematically comparatively tractable, yet computationally powerful nonlinear recurrent network approach that can reproduce a wide range of nonlinear dynamical phenomena [75-78]. Given its semi-analytical nature, the present algorithm runs reasonably fast (and starting it from a number of different initializations to approach a globally optimal solution is computationally very feasible). However, its mathematical properties (among them, issues of convergence/monotonicity, local maxima/uniqueness and existence of solutions, and identifiability [of dynamics]) certainly require further illumination and may lead to further algorithmic improvements.

*Mechanisms of working memory*

Although the primary focus of this work was to develop and evaluate a state space framework for PLRNNs, some discussion of the applicational example chosen here, working memory, is in order. Working memory is generally defined as the ability to actively hold an item in memory, in the absence of guiding external input, for short-term reference in subsequent choice situations [79]. Various neural mechanisms have been proposed to underlie this cognitive capacity, most prominently multi-stable neural networks which retain short-term memory items by switching into one of several stimulus-selective attractor states (e.g. [24, 25, 28]). These attractors usually represent fixed points in the firing rates, with assemblies of recurrently coupled stimulus-selective cells exhibiting high rates while those cells not coding for the present stimulus in short-term memory remaining at a spontaneous low-rate base level. These models were inspired by the physiological observation of 'delay-active' cells [80-82], that is cells that switch into a high-rate state during the delay periods of working memory tasks, and back to a low-rate state after completion of a trial, similar to the 'delay-active' latent states observed in Fig. 8. Nakahara & Doya [83] were among the first to point out, however, that, for working memory, it may be completely sufficient (or even advantageous) to tune the system close to a bifurcation point where the dynamics becomes very slow (see also [1]), and true multi-stability may not be required. This is supported by the present observation that most of the estimated PLRNN models had fixed points with eigenvalues close to 1 but were not truly bi- or multi-stable (cf. Fig. 9), yet this was sufficient to account for maintenance of stimulus-selectivity throughout the 10 s delay of the present task (cf. Fig. 8) and for experimental observations (cf. Fig. 7). Recently, a number of other mechanisms for supporting working memory, however, including sequential activation of cell populations [84] or synaptic mechanisms [85] have been discussed. Thus, the neural mechanisms of working memory remain an active research area to which statistical model estimation approaches as developed here may significantly contribute, but too broad a topic in its own right to be covered in more depth by this mainly methodological work.

**Models and Methods**

*Expectation-Maximization Algorithm: State estimation*

As with most previous work on estimation in (neural) state space models [16, 18, 19, 22], we use the Expectation-Maximization (EM) framework for obtaining estimates of both the model parameters and the underlying latent state path. Due to the piecewise-linear nature of model (1), however, the conditional latent state path density p(**Z**|**X**) is a high-dimensional 'mixture' of partial Gaussians, with the number of integrations to perform to obtain moments of p(**Z**|**X**) scaling as $2^{T \times M}$. Although analytically accessible, this will be computationally prohibitive for almost all cases of interest. Our approach therefore focuses on a computationally reasonably efficient way of searching for the mode (maximum a-posteriori, MAP estimate) of p(**Z**|**X**) which was found to be in good agreement with E(**Z**|**X**) in most cases. Covariances were then approximated locally around the MAP estimate.

More specifically, the EM algorithm maximizes the expected log-likelihood of the joint distribution p(**X**,**Z**) as a lower bound on $\log p(\mathbf{X} | \Xi)$ [23], where $\Xi = \{\boldsymbol{\mu}_0, \mathbf{A}, \mathbf{W}, \boldsymbol{\Sigma}, \mathbf{B}, \boldsymbol{\Gamma}\}$ denotes the set of to-be-optimized-for parameters (note that we dropped the thresholds θ from this for now):

$$
\begin{aligned}
Q(\Xi, \mathbf{Z}) &:= E[\log p(\mathbf{Z}, \mathbf{X} | \Xi)] \\
&= E\left[-\frac{1}{2}(\mathbf{z}_1 - \boldsymbol{\mu}_0 - \mathbf{s}_1)^T \boldsymbol{\Sigma}^{-1}(\mathbf{z}_1 - \boldsymbol{\mu}_0 - \mathbf{s}_1)\right] \\
&\quad + E\left[-\frac{1}{2}\sum_{t=2}^{T}(\mathbf{z}_t - \mathbf{A}\mathbf{z}_{t-1} - \mathbf{W}\phi(\mathbf{z}_{t-1}) - \mathbf{s}_t)^T \boldsymbol{\Sigma}^{-1}(\mathbf{z}_t - \mathbf{A}\mathbf{z}_{t-1} - \mathbf{W}\phi(\mathbf{z}_{t-1}) - \mathbf{s}_t)\right] \\
&\quad + E\left[-\frac{1}{2}\sum_{t=1}^{T}(\mathbf{x}_t - \mathbf{B}\phi(\mathbf{z}_t))^T \boldsymbol{\Gamma}^{-1}(\mathbf{x}_t - \mathbf{B}\phi(\mathbf{z}_t))\right] - \frac{T}{2}(\log|\boldsymbol{\Sigma}| + \log|\boldsymbol{\Gamma}|).
\end{aligned}
$$

(5)

For state estimation (*E*-step), if $\phi$ were a linear function, obtaining $E(\mathbf{Z} | \mathbf{X}, \Xi)$ would be equivalent to maximizing the argument of the expectancy in (5) w.r.t. **Z**, i.e., $\mathrm{E}[\mathbf{Z} | \mathbf{X}, \Xi] \equiv \arg\max_{\mathbf{Z}}[\log p(\mathbf{Z}, \mathbf{X} | \Xi)]$ (see [12]; see also [86]). This is because for a Gaussian mean and mode coincide. In our case, p(**X**,**Z**) is piecewise Gaussian, and we still take the approach (suggested in [12]) of maximizing $\log p(\mathbf{Z}, \mathbf{X} | \Xi)$ directly w.r.t. **Z** (essentially a Laplace approximation of $p(\mathbf{X} | \Xi)$ where we neglect the Hessian which is constant around the maximizer; cf. [12,37]).

Let $\Omega(t) \subseteq \{1...M\}$ be the set of all indices of the units for which we have $z_{mt} \leq \theta_m$ at time *t*, and $\mathbf{W}_{\Omega(t)}$ and $\mathbf{B}_{\Omega(t)}$ the matrices **W** and **B**, respectively, with all columns with indices $\in \Omega(t)$ set to 0. The state estimation problem can then be formulated as

(6)     *maximize*

$$Q_\Omega^*(\mathbf{Z}) := -\frac{1}{2}(\mathbf{z}_1 - \boldsymbol{\mu}_0 - \mathbf{s}_1)^T \boldsymbol{\Sigma}^{-1}(\mathbf{z}_1 - \boldsymbol{\mu}_0 - \mathbf{s}_1)$$

$$-\frac{1}{2}\sum_{t=2}^T [\mathbf{z}_t - (\mathbf{A} + \mathbf{W}_{\Omega(t-1)})\mathbf{z}_{t-1} + \mathbf{W}_{\Omega(t-1)}\boldsymbol{\theta} - \mathbf{s}_t]^T \boldsymbol{\Sigma}^{-1}[\mathbf{z}_t - (\mathbf{A} + \mathbf{W}_{\Omega(t-1)})\mathbf{z}_{t-1} + \mathbf{W}_{\Omega(t-1)}\boldsymbol{\theta} - \mathbf{s}_t]$$

$$-\frac{1}{2}\sum_{t=1}^T (\mathbf{x}_t - \mathbf{B}_{\Omega(t)}\mathbf{z}_t + \mathbf{B}_{\Omega(t)}\boldsymbol{\theta})^T \boldsymbol{\Gamma}^{-1}(\mathbf{x}_t - \mathbf{B}_{\Omega(t)}\mathbf{z}_t + \mathbf{B}_{\Omega(t)}\boldsymbol{\theta})$$

w.r.t. $(\Omega, \mathbf{Z})$, subject to $z_{tm} \leq \theta_m \;\forall t, m \in \Omega(t)$ AND $z_{tm} > \theta_m \;\forall t, m \notin \Omega(t)$

Let us concatenate all state variables into one long column vector, $\mathbf{z} = (\mathbf{z}_1, ..., \mathbf{z}_T) = (z_{11}...z_{mt}...z_{MT})^T$, and unwrap the sums across time into large, block-banded *MT×MT* matrices (see [12, 71]) in which we combine all terms quadratic or linear in **z**, or $\phi(\mathbf{z})$, respectively. Further, define $\mathbf{d}_\Omega$ as the binary (*MT*×1) indicator vector which has 1s everywhere except for the entries with indices $\in \Omega \subseteq \{1...MT\}$ which are set to 0, and let $\mathbf{D}_\Omega := diag(\mathbf{d}_\Omega)$ the *MT*×*MT* diagonal matrix formed from $\mathbf{d}_\Omega$. Let $\boldsymbol{\Theta} := (\boldsymbol{\theta}, \boldsymbol{\theta}, ..., \boldsymbol{\theta})_{(MT \times 1)}$, and $\boldsymbol{\Theta}^{+M}$ the same vector shifted downward by *M* positions, with the first *M* entries set to 0. One may then rewrite $Q_\Omega^*(\mathbf{Z})$ in the form

(7)
$$Q_\Omega^*(\mathbf{Z}) = -\frac{1}{2}\Big[\mathbf{z}^T(\mathbf{U}_0 + \mathbf{D}_\Omega \mathbf{U}_1 + \mathbf{U}_1^T \mathbf{D}_\Omega + \mathbf{D}_\Omega \mathbf{U}_2 \mathbf{D}_\Omega)\mathbf{z}$$
$$-\mathbf{z}^T(\mathbf{v}_0 + \mathbf{D}_\Omega \mathbf{v}_1 + \mathbf{V}_2 diag[\mathbf{d}_\Omega^{+M}]\boldsymbol{\Theta}^{+M} + \mathbf{V}_3 \mathbf{D}_\Omega \boldsymbol{\Theta} + \mathbf{D}_\Omega \mathbf{V}_4 \mathbf{D}_\Omega \boldsymbol{\Theta})$$
$$-(\mathbf{v}_0 + \mathbf{D}_\Omega \mathbf{v}_1 + \mathbf{V}_2 diag[\mathbf{d}_\Omega^{+M}]\boldsymbol{\Theta}^{+M} + \mathbf{V}_3 \mathbf{D}_\Omega \boldsymbol{\Theta} + \mathbf{D}_\Omega \mathbf{V}_4 \mathbf{D}_\Omega \boldsymbol{\Theta})^T \mathbf{z}\Big] + const.$$

The *MT*×*MT* matrices $\mathbf{U}_{\{0...2\}}$ separate product terms that do not involve $\phi(\mathbf{z})$ ($\mathbf{U}_0$), involve multiplication by $\phi(\mathbf{z})$ only from the left-hand or right-hand side ($\mathbf{U}_1$), or from both sides ($\mathbf{U}_2$). Likewise, for the terms linear in **z**, vector and matrix terms were separated that involved $z_{mt}$ or $\theta_m$ conditional on $z_{mt} > \theta_m$ (please see the provided MatLab code for the exact composition of these matrices). For now, the important point is that we have $2^{M \times T}$ different quadratic equations, depending on the bits on and off in the binary vector $\mathbf{d}_\Omega$. Consequently, to obtain the MAP estimator for **z**, in theory, one may consider all $2^{M \times T}$ different settings for $\mathbf{d}_\Omega$, for each solve the linear equations implied by $\partial Q_\Omega^*(\mathbf{Z})/\partial \mathbf{Z} = 0$, and select among those for which the solution $\mathbf{z}_*$ is consistent with the considered set $\Omega$ the one which produces the largest value $Q_\Omega^*(\mathbf{z}_*)$.

In practice, this is generally not feasible. Various solution methods for piecewise linear equations have been suggested in the mathematical programming literature in the past [87, 88]. For instance, some piecewise linear problems may be recast as a linear complementarity problem [89], but the pivoting methods often used to solve it work (numerically) well only for smaller scale settings [38]. Here we therefore settled on a similar, simple Newton-type iteration scheme as proposed in Brugnano & Casulli [38]. Specifically, if we denote by $\mathbf{z}_*(\Omega)$ the solution to eq. 7 obtained with the set of constraints $\Omega$ active, the present scheme initializes with a random drawing of the $\{z_{mt}\}$, sets the components of $\mathbf{d}_\Omega$ for which $z_{mt} > \theta_m$ to 1 and all others to 0, and then keeps on alternating between (1) solving $\partial Q_\Omega^*(\mathbf{Z})/\partial \mathbf{Z} = 0$ for $\mathbf{z}_*(\Omega)$ and (2) flipping the bits in $\mathbf{d}_\Omega$ for which $\text{sgn}[2d_\Omega^{(k)} - 1] \neq \text{sgn}[z_{*k}(\Omega) - \theta_k]$, that is, for which the components of the vector

(8)    $\mathbf{c} := (2\mathbf{d}_\Omega - \mathbf{1})^T \circ (\mathbf{\theta} - \mathbf{z}_*(\Omega))$

are positive, until the solution to $\partial Q_\Omega^*(\mathbf{Z})/\partial \mathbf{Z} = 0$ is consistent with set $\Omega$ (i.e., $\mathbf{c} \leq \mathbf{0}$).

For the problem as formulated in Brugnano & Casulli [38], these authors proved that such a solution always exists, and that the algorithm will always terminate after a finite (usually low) number of steps, given certain assumptions and provided the matrix that multiplies with the states **z** in $\partial Q_\Omega^*(\mathbf{Z})/\partial \mathbf{Z} = 0$ (i.e., the Hessian of $Q_\Omega^*(\mathbf{z}_*)$), fulfills certain conditions (Stieltjes-type; see [38] for details). This will usually *not* be the case for the present system; although the Hessian of $Q_\Omega^*(\mathbf{z}_*)$ will be symmetric and positive-definite (with proper parameter settings), its off-diagonal elements may be either larger or smaller than 0. Moreover, for the problem considered here, all elements of the Hessian in (7) depend on $\Omega$, while in [38] this is only the case for the on-diagonal elements (i.e., in [38] $\mathbf{D}_\Omega$ enters the Hessian only in additive, not multiplicative form as here). For these reasons, the Newton-type algorithm outlined above may not always converge to an exact solution (if one exists in this case) but may eventually cycle among non-solution configurations, or may not even always increase $Q(\mathbf{Z})$ (i.e., eq. 5!). To bypass this, the algorithm was always terminated if one of the following three conditions was met: (i) A solution to $\partial Q_\Omega^*(\mathbf{Z})/\partial \mathbf{Z} = 0$ consistent with $\Omega$ was encountered; (ii) a previously probed set $\Omega$ was revisited; (iii) the constraint violation error defined by $\|\mathbf{c}_+\|_1$, the $l_1$ norm of the positive part of **c** defined in eq. 8, went up beyond a pre-specified tolerance level. With these modifications, we found that the algorithm would usually terminate after only a few iterations (<10 for the examined toy examples) and yield approximate solutions with only a few constraints still violated (<3% for the toy examples). For these elements *k* of **z** for which the constraints are still violated, that is for which $c_k > 0$ in eq. 8, one may explicitly enforce the constraints by setting the violating states $\mathbf{z}_{\{k\}} = \mathbf{\theta}_{\{k\}}$, but either way it was found that even these approximate (and potentially only locally optimal) solutions were generally (for the problems studied) in sufficiently good agreement with E(**Z**|**X**).

In the case of full EM iterations (with the parameters unknown as well), it appeared that flipping violated constraints in $\mathbf{d}_\Omega$ one by one may often (for the scenarios studied here) improve overall performance, in the sense of yielding higher-likelihood solutions and less numerical problems (although it may leave more constraints violated in the end). Hence, this scheme was adopted here for the full EM, that is only the single bit $k^*$ corresponding to the maximum element of vector **c** in eq. 8 was inverted on each iteration (the one with the largest wrong-side deviation from $\mathbf{\theta}$). In general, however, the resultant slow-down in the algorithm may not always be worth the performance gains; or a mixture of methods, with $d_{k^*}^{l+1} = 1 - d_{k^*}^l$ with $k^* := \arg\max_k \{c_k > 0\}$ early on, and $\mathbf{d}_{\{k\}}^{l+1} = \mathbf{1} - \mathbf{d}_{\{k\}}^l \, \forall k : c_k > 0$ during later iterations, may be considered.

Once a (local) maximum $\mathbf{z}^{\max}$ has been obtained, the (local) covariances may be read off from the inverse negative Hessian at $\mathbf{z}^{\max}$, i.e. the elements of

(9) $\quad \mathbf{V} := (\mathbf{U}_0 + \mathbf{D}_\Omega \mathbf{U}_1 + \mathbf{U}_1^T \mathbf{D}_\Omega + \mathbf{D}_\Omega \mathbf{U}_2 \mathbf{D}_\Omega)^{-1}.$

We then use these covariance estimates to obtain (estimates of) $\mathrm{E}[\phi(\mathbf{z})]$, $\mathrm{E}[\mathbf{z}\phi(\mathbf{z}^T)]$, and $\mathrm{E}[\phi(\mathbf{z})\phi(\mathbf{z}^T)]$, as required for the maximization step. Denoting by $F(\lambda;\mu,\sigma^2) := \int_\lambda^\infty N(x;\mu,\sigma^2)dx$ the complementary cumulative Gaussian, to ease subsequent derivations, let us introduce the following notation:

(10) $\quad N_k := N(\theta_k; z_k^{\max}, \sigma_k^2), \quad F_k := F(\theta_k; z_k^{\max}, \sigma_k^2), \quad \sigma_{kl}^2 := \mathrm{cov}(z_k^{\max}, z_l^{\max}) \approx v_{kl}.$

The elements of the expectancy vectors and matrices above are computed as

(11) $\quad \mathrm{E}[\phi(z_k)] = \sigma_k^2 N_k + (z_k^{\max} - \theta_k) F_k,$

$\mathrm{E}[\phi(z_k)^2] = ([z_k^{\max}]^2 + \sigma_k^2 + \theta_k^2 - 2\theta_k z_k^{\max}) F_k + (z_k^{\max} - \theta_k)\sigma_k^2 N_k,$

$\mathrm{E}[z_k \phi(z_l)] = (\sigma_{kl}^2 - \theta_l z_k^{\max} + z_k^{\max} z_l^{\max}) F_l + z_k^{\max} \sigma_l^2 N_l.$

The terms $\mathrm{E}[\phi(z_k)\phi(z_l)]$, for $k \neq l$, are more tedious, and cannot be (to my knowledge and insight) computed exactly (analytically), so we develop them in a bit more detail here:

(12)
$$\mathrm{E}[\phi(z_k)\phi(z_l)] = \int_{\theta_k}^\infty \int_{\theta_l}^\infty p(z_k, z_l)(z_k - \theta_k)(z_l - \theta_l) dz_k dz_l$$
$$= \int_{\theta_k}^\infty \int_{\theta_l}^\infty p(z_k, z_l) z_k z_l dz_k dz_l - \theta_k \int_{\theta_k}^\infty \int_{\theta_l}^\infty p(z_k, z_l) z_l dz_k dz_l - \theta_l \int_{\theta_k}^\infty \int_{\theta_l}^\infty p(z_k, z_l) z_k dz_k dz_l$$
$$+ \theta_k \theta_l \int_{\theta_k}^\infty \int_{\theta_l}^\infty p(z_k, z_l) dz_k dz_l$$

The last term is just a (complementary) cumulative bivariate Gaussian evaluated with parameters specified through the MAP solution $(\mathbf{z}^{\max}, \mathbf{V})$ (and multiplied by the thresholds). The first term we may rewrite as follows:

$$\int_{\theta_k}^\infty \int_{\theta_l}^\infty p(z_k, z_l) z_k z_l dz_k dz_l = \int_{\theta_k}^\infty p(z_k) z_k \int_{\theta_l}^\infty p(z_l | z_k) z_l dz_k dz_l$$
$$= \int_{\theta_k}^\infty p(z_k) z_k \left[ N(\theta_l; \mu_{l|k}, \lambda_l^{-1}) + \mu_{l|k}\left(1 - \int_{-\infty}^{\theta_l} N(z_l; \mu_{l|k}, \lambda_{lk}^{-1}) dz_l \right) \right] dz_k$$

(13) $\quad$ where $\quad \mu_{l|k} := z_l^{\max} - \lambda_l^{-1} \lambda_{lk}(z_k - z_k^{\max})$

$\lambda_l := \sigma_l^2 / (\sigma_k^2 \sigma_l^2 - \sigma_{kl}^4)$

$\lambda_{lk} := -\sigma_{kl}^2 / (\sigma_k^2 \sigma_l^2 - \sigma_{kl}^4)$

These are just standard results one can derive by the reverse chain rule for integration, with the $\lambda$'s the elements of the inverse bivariate ($k,l$)-covariance matrix. Note that if the variable $z_k$ were removed from the first integrand in eq. 13, i.e. as in the second term in eq. 12, all terms in eq. 13 would just come down to uni- or bivariate Gaussians (times some factor) or a univariate Gaussian expectancy value, respectively. Noting this, one obtains for the second (and correspondingly for the third) term in eq. 12:

(14)
$$\theta_k \int_{\theta_k}^{\infty}\int_{\theta_l}^{\infty} p(z_k,z_l)z_l dz_k dz_l = \theta_k \lambda_k N_l F(\theta_k;\mu_{lk},\lambda_l^{-1}) + \theta_k(z_l^{\max}F_k + \sigma_{kl}^2 N_k)F(\theta_l;z_l^{\max},\lambda_k^{-1})$$
$$\text{with } \mu_{kl} := z_l^{\max} + \sigma_{kl}^2/\sigma_k^2(\theta_k - z_k^{\max})$$

The problematic bit is the product term $\int_{\theta_k}^{\infty} p(z_k)z_k \mu_{l|k} \int_{-\infty}^{\theta_l} N(z_l;\mu_{l|k},\lambda_{lk}^{-1})dz_l dz_k$ in eq. 13, which we resolve by making the approximation $\mu_{l|k} \approx \mu_l = z_l^{\max}$. This way we have for the first term in eq. 12:

(15)
$$\int_{\theta_k}^{\infty}\int_{\theta_l}^{\infty} p(z_k,z_l)z_k z_l dz_k dz_l \approx \lambda_k N_l \left[\lambda_l^{-1}N(\theta_k;\mu_{lk},\lambda_l^{-1}) + \mu_{lk}F(\theta_k;\mu_{lk},\lambda_l^{-1})\right]$$
$$+ \left[(\sigma_k^2 z_l^{\max} - \sigma_{kl}^2 z_k^{\max})N_k + (z_k^{\max}z_l^{\max} + \sigma_{kl}^2)F_k\right]F(\theta_l;\mu_{lk},\lambda_k^{-1})$$

Putting (13)-(15) together with the bivariate cumulative Gaussian yields an analytical approximation to eq. 12 that can be computed based on the quantities obtained from the MAP estimate $(\mathbf{z}^{\max}, \mathbf{V})$.

*Expectation-Maximization Algorithm: Parameter estimation*

Once we have estimates for $E[\mathbf{z}]$, $E[\mathbf{z}\mathbf{z}^T]$, $E[\phi(\mathbf{z})]$, $E[\mathbf{z}\phi(\mathbf{z}^T)]$, and $E[\phi(\mathbf{z})\phi(\mathbf{z}^T)]$, the maximization step is standard and straightforward, so for convenience we just state the results here, using the notation

(16)
$$\mathbf{E}_{1,\Delta} := \sum_{t=1}^{T-\Delta} E[\phi(\mathbf{z}_t)\phi(\mathbf{z}_t^T)] \quad, \quad \mathbf{E}_2 := \sum_{t=2}^{T} E[\mathbf{z}_t\mathbf{z}_{t-1}^T], \quad \mathbf{E}_{3,\Delta} := \sum_{t=1+\Delta}^{T-1+\Delta} E[\mathbf{z}_t\mathbf{z}_t^T],$$
$$\mathbf{E}_4 := \sum_{t=1}^{T-1} E[\phi(\mathbf{z}_t)\mathbf{z}_t^T] \quad, \quad \mathbf{E}_5 := \sum_{t=2}^{T} E[\mathbf{z}_t\phi(\mathbf{z}_{t-1}^T)],$$
$$\mathbf{F}_1 := \sum_{t=1}^{T} \mathbf{x}_t E[\phi(\mathbf{z}_t^T)] \quad, \quad \mathbf{F}_2 := \sum_{t=1}^{T} \mathbf{x}_t \mathbf{x}_t^T \quad, \quad \mathbf{F}_3 := \sum_{t=2}^{T} \mathbf{s}_t E[\mathbf{z}_{t-1}^T],$$
$$\mathbf{F}_4 := \sum_{t=2}^{T} \mathbf{s}_t E[\phi(\mathbf{z}_{t-1}^T)] \quad, \quad \mathbf{F}_5 := \sum_{t=1}^{T} E[\mathbf{z}_t]\mathbf{s}_t^T \quad, \quad \mathbf{F}_6 := \sum_{t=1}^{T} \mathbf{s}_t \mathbf{s}_t^T$$

With these expectancy sums defined, one has

(17a)   $\mathbf{B} = \mathbf{F}_1 \mathbf{E}_{1,0}^{-1}$

(17b)   $\mathbf{\Gamma} = \frac{1}{T}(\mathbf{F}_2 - \mathbf{F}_1\mathbf{B}^T - \mathbf{B}\mathbf{F}_1^T + \mathbf{B}\mathbf{E}_{1,0}^T\mathbf{B}^T) \circ \mathbf{I}$

(17c)   $\mathbf{\mu}_0 = E[\mathbf{z}_1] - \mathbf{s}_1$

(17d)   $\mathbf{A} = [(\mathbf{E}_2 - \mathbf{W}\mathbf{E}_4 - \mathbf{F}_3) \circ \mathbf{I}][\mathbf{E}_{3,0} \circ \mathbf{I}]^{-1}$

(17e)

$$\mathbf{\Sigma} = \frac{1}{T}\Big[\text{var}(\mathbf{z}_1) + \mathbf{\mu}_0\mathbf{s}_1^T + \mathbf{s}_1\mathbf{\mu}_0^T + \mathbf{E}_{3,1}^T - \mathbf{F}_5 - \mathbf{F}_5^T + \mathbf{F}_6 + (\mathbf{F}_3 - \mathbf{E}_2)\mathbf{A}^T + \mathbf{A}(\mathbf{F}_3^T - \mathbf{E}_2^T) + \mathbf{A}\mathbf{E}_{3,0}^T\mathbf{A}^T$$
$$+ (\mathbf{F}_4 - \mathbf{E}_5)\mathbf{W}^T + \mathbf{W}(\mathbf{F}_4^T - \mathbf{E}_5^T) + \mathbf{W}\mathbf{E}_1^T\mathbf{W}^T + \mathbf{A}\mathbf{E}_4^T\mathbf{W}^T + \mathbf{W}\mathbf{E}_4\mathbf{A}^T\Big] \circ \mathbf{I}$$

Note that to avoid redundancy in the parameters, here we usually fixed $\Sigma = \mathbf{I} \cdot 10^{-2}$.

For **W**, since we assumed this matrix to have an off-diagonal structure (i.e., with zeros on the diagonal), we solve for each row of **W** separately:

(17f)
$$\mathbf{P}^{(0)} := (\mathbf{E}_{3,0} \circ \mathbf{I})^{-1} \mathbf{E}_4^T$$
$$\mathbf{P}^{(1)} := \mathbf{E}_5 - [(\mathbf{E}_2 - \mathbf{F}_3) \circ \mathbf{I}] \mathbf{P}^{(0)} - \mathbf{F}_4$$
$$\forall m \in \{1...M\} : \mathbf{W}_{m,\{1:M\}\backslash m} = \mathbf{P}^{(1)}_{m,\{1:M\}\backslash m} ([\mathbf{E}_1 - \mathbf{E}_{4,\bullet m} \mathbf{P}^{(0)}_{m\bullet}]_{\{1:M\}\backslash m, \{1:M\}\backslash m})^{-1}$$

where the subscripts indicate the matrix elements to be pulled out, with the subscript dot denoting all elements of the corresponding column or row (e.g., '•m' takes the $m^{th}$ column of that matrix).

Starting from a number of different random parameter initializations, the E- and M-steps are alternated until the log-likelihood ratio falls below a predefined tolerance level (while still increasing) or a preset maximum number of allowed iterations are exceeded. For reasons mentioned in the Results, sometimes it can actually happen that the log-likelihood ratio temporarily decreases, in which case the iterations are continued. If $(N-M)^2 \geq N+M$, factor analysis may be used to derive initial estimates for the latent states and observation parameters in (3) [23], although this was not attempted here. For further implementational details see the MatLab code provided at www.zi-mannheim.de/en/research/departments-research-groups-institutes/theor-neuroscience-e.html [**upon publication**].

*Particle filter*

To validate the approximations from our semi-analytical procedure developed above, a bootstrap particle filter as given in Durbin & Koopman [22] was implemented. In bootstrap particle filtering, the state posterior distribution at time *t*,

(18)
$$p_\Xi(\mathbf{z}_t | \mathbf{x}_1,...,\mathbf{x}_t) = \frac{p_\Xi(\mathbf{x}_t | \mathbf{z}_t) p_\Xi(\mathbf{z}_t | \mathbf{x}_1,...,\mathbf{x}_{t-1})}{p_\Xi(\mathbf{x}_t | \mathbf{x}_1,...,\mathbf{x}_{t-1})}$$
$$= \frac{p_\Xi(\mathbf{x}_t | \mathbf{z}_t) \int_{\mathbf{z}_{t-1}} p_\Xi(\mathbf{z}_t | \mathbf{z}_{t-1}) p_\Xi(\mathbf{z}_{t-1} | \mathbf{x}_1,...,\mathbf{x}_{t-1}) d\mathbf{z}_{t-1}}{p_\Xi(\mathbf{x}_t | \mathbf{x}_1,...,\mathbf{x}_{t-1})}$$

is numerically approximated through a set of 'particles' (samples) $\{\mathbf{z}_t^{(1)},...,\mathbf{z}_t^{(K)}\}$, drawn from $p_\Xi(\mathbf{z}_t | \mathbf{x}_1,...,\mathbf{x}_{t-1})$, together with a set of normalized weights $\{w_t^{(1)},...,w_t^{(K)}\}$,

$w_t^{(r)} := p_\Xi(\mathbf{x}_t | \mathbf{z}_t^{(r)}) \left(\sum_{k=1}^K p_\Xi(\mathbf{x}_t | \mathbf{z}_t^{(k)})\right)^{-1}$. Based on this representation, moments of $p_\Xi(\mathbf{z}_t | \mathbf{x}_{1:t})$ and $p_\Xi(\phi(\mathbf{z}_t) | \mathbf{x}_{1:t})$ can be easily obtained by evaluating $\phi$ (or any other function of **z**) on the set of samples $\{\mathbf{z}_t^{(r)}\}$ and summing the outcomes weighted with their respective normalized observation likelihoods $\{w_t^{(r)}\}$. A new set of samples $\{\mathbf{z}_{t+1}^{(r)}\}$ for *t*+1 is then generated by first drawing *K* times from $\{\mathbf{z}_t^{(k)}\}$ with replacement according to the weights $\{w_t^{(k)}\}$, and then drawing *K* new samples according to the transition probabilities $p_\Xi(\mathbf{z}_{t+1}^{(k)} | \mathbf{z}_t^{(k)})$ (thus approximating the integral in eq. 18). Here we used $K=10^4$ samples. Note that this numerical sampling scheme, like a Kalman filter, but

unlike the procedure outlined above, only implements the filtering step (i.e., yields $p_\Xi(\mathbf{z}_t \mid \mathbf{x}_{1:t})$, not $p_\Xi(\mathbf{z}_t \mid \mathbf{x}_{1:T})$). On the other hand, it gives (weakly) consistent (asymptotically unbiased; [90, 91]) estimates of all expectancies across this distribution, that is, it does not rely on the type of approximations and locally optimal solutions of our semi-analytical approach that almost inevitably will come with some bias (since, among other factors, the mode would usually deviate from the mean by some amount for the present model).

*Experimental data sets*

Details of the experimental task and electrophysiological data sets used here could be found in Hyman et al. [35]. Briefly, rats had to alternate between left and right lever presses in a Skinner box to obtain a food reward dispensed on correct choices, with a $\geq 10$ s delay enforced between consecutive lever presses. While the levers were located on one side of the Skinner box, animals had to perform a nosepoke on the opposite side of the box in between lever presses for initiating the delay period, to discourage them from developing an external coding strategy (e.g., through maintenance of body posture during the delay). While animals were performing the task, multiple single units were recorded with a set of 16 tetrodes implanted bilaterally into the anterior cingulate cortex (ACC, a subdivision of rat prefrontal cortex). For the present analyses, a data set from only one of the four rats recorded on this task was selected for the present exemplary purposes, namely the one where the clearest single unit traces of delay activity were observed in the first place. This data set consisted of 30 simultaneously recorded units, of which the 19 units with spiking rates >1 Hz were retained, on 14 correct trials (only correct response trials were analyzed). The trials had variable length, but were all cut down to the same length of 14 s, including 2 s of pre-nosepoke, 5 s extending into the delay from the nosepoke, 5 s preceding the next lever press, and 2 s of post-response phase (note that this may imply temporal gaps in the middle of the delay on some trials, which were ignored here for convenience). All spike trains were convolved with Gaussian kernels (see, e.g., [8, 46, 92]), with the kernel standard deviation set individually for each unit to one half of its mean interspike-interval. Note that this also brings the observed series into tighter agreement with the Gaussian assumptions of the observation model, eq. 3. Finally, the spike time series were binned into 500 ms bins (corresponding roughly to the inverse of the overall [across all 30 recorded cells] average neural firing rates of $\approx 2.2$ Hz), which resulted in 14 trials of 28 time bins each submitted to the estimation process. As indicated in the section '*State space model*', a trial-unique initial state mean $\boldsymbol{\mu}_k, k = 1\ldots 14$, was assumed for each of the 14 temporally segregated trials.

**Acknowledgements**

I thank Dr. Georgia Koppe for her feedback on this manuscript, and Drs. James Hyman and Jeremy Seamans for lending me their in-vivo electrophysiological recordings from rat ACC as an analysis testbed.

**Funding statement**

This work was funded through two grants from the German Research Foundation (DFG, Du 354/8-1, and within the Collaborative Research Center 1134) to the author, and by the German Ministry for Education and Research (BMBF, 01ZX1314G) within the e:Med program.

**Figure Legends**

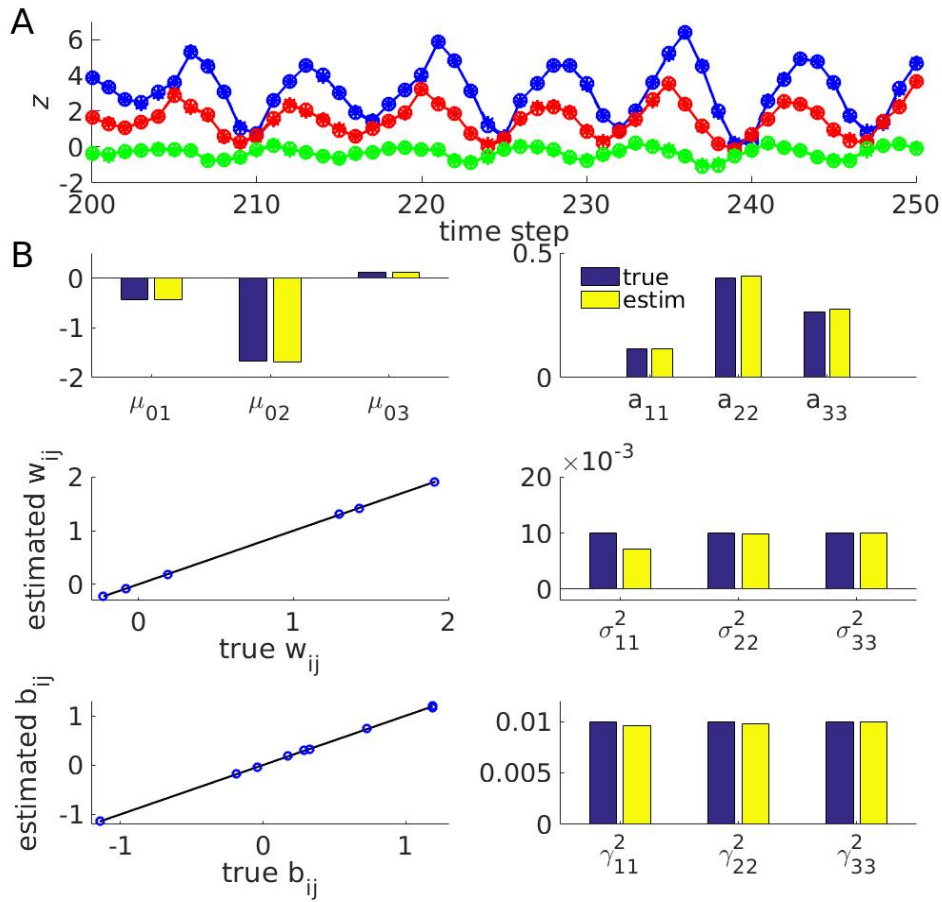

**Fig. 1.** State and parameter estimates for nonlinear cycle example. (A) True (solid-circle lines) and estimated (dashed-star lines) states over some periods of the simulated limit cycle generated by a 3-state PLRNN when true parameters were provided (for this example, $\boldsymbol{\theta} \approx (0.86, 0.09, -0.85)$; all other parameters as in B, see also provided Matlab file 'PLRNNoscParam.mat'). 'True states' refers to the actual states from which the observations **X** were generated. Inputs of $s_{it} = 1$ were provided to units *i*=1 and *i*=2 on time steps 1 and 10 of each cycle, respectively. (B) True and estimated model parameters for (from top-left to bottom-right) $\boldsymbol{\mu}_0, \mathbf{A}, \mathbf{W}, \boldsymbol{\Sigma}, \mathbf{B}, \boldsymbol{\Gamma}$, when true states (but not their higher-order moments) were provided. Bisectrix lines (black) indicate identity.

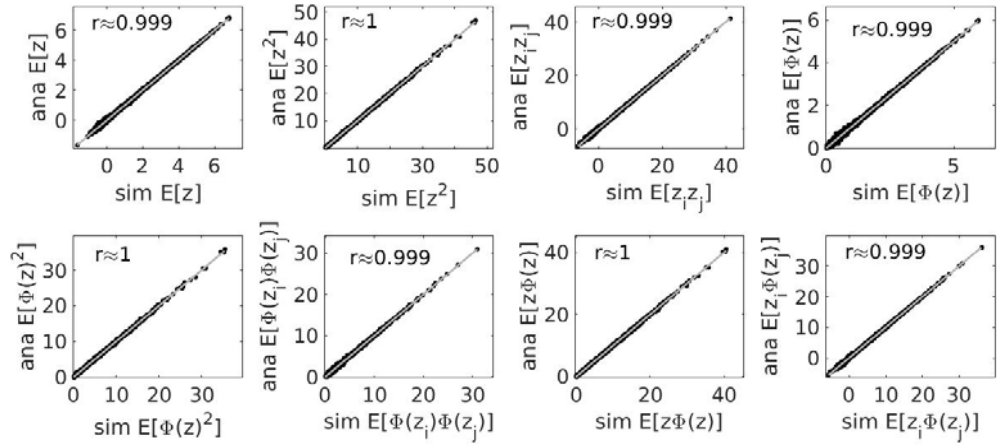

**Fig. 2.** Agreement between simulated (*x*-axes) and semi-analytical (*y*-axes) solutions for state expectancies for the model from Fig. 1 across all three state variables and *T*=750 time steps. Here, $\phi(z_i) := \max\{0, z_i - \theta_i\}$ is the PL activation function. Simulated state paths and their moments were generated using a bootstrap particle filter with $10^4$ particles. Bisectrix lines in gray indicate identity.

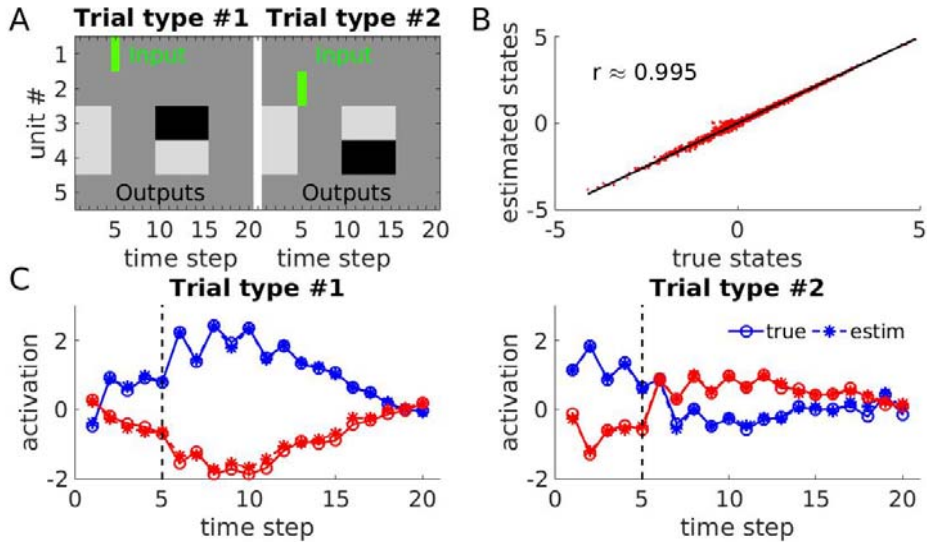

**Fig. 3.** State estimation for 'working memory' example when true parameters were provided. (A) Setup of the simulated working memory task: Stimulus inputs (green bars, $s_{it}=1$, and 0 otherwise) and requested outputs (black = 1, light-gray = 0, dark-grey = no output required) across the 20 time points of a working memory trial (with two different trial types) for the 5 PLRNN units. (B) Correlation between estimated and true states (i.e., those from which the observations **X** were generated) across all five state variables and *T*=800 time steps. Bisectrix in black. (C) True (circle-solid lines) and estimated (star-dashed lines) states for output units #3 (blue) and #4 (red) when $s_{15}=1$ (left) or $s_{25}=1$ (right) for single example trials. Note that although working memory PLRNNs may, in principle, be explicitly designed, here a 5-state PLRNN was first trained by conventional gradient descent (real-time recurrent-learning; [39]) to perform the task in A, to yield more 'natural' and less uniform ground truth states and parameters. Here, all $\theta_i = 0$ (implying that there can only be one stable fixed point). See Matlab file 'PLRNNwmParam.mat' and Fig. 4 for details on parameters.

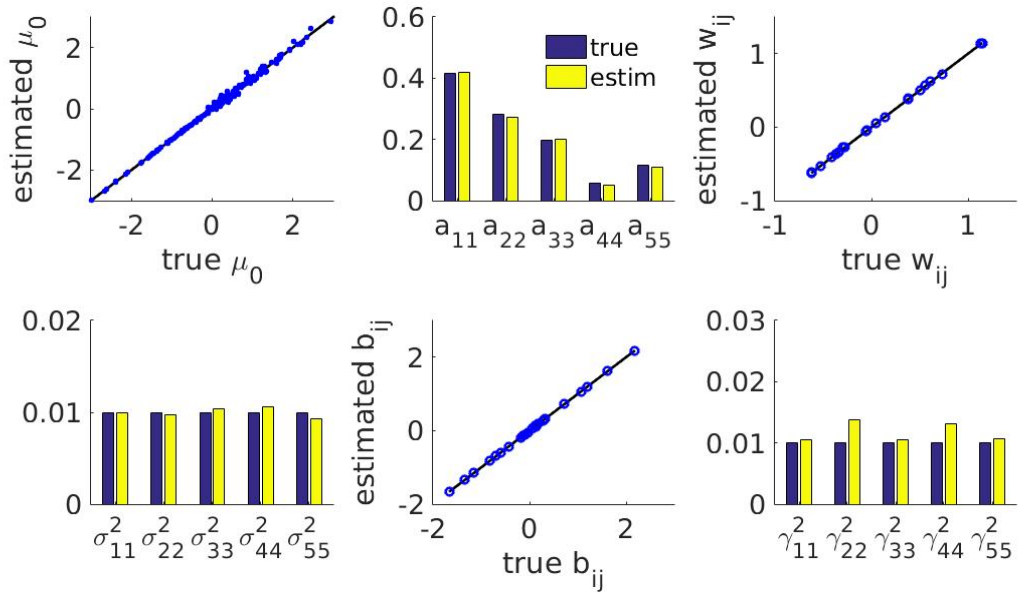

**Fig. 4.** True and estimated parameters for the working memory PLRNN (cf. Fig. 3) when true states were provided. From top-left to bottom-right, estimates for: $\boldsymbol{\mu}_0, \mathbf{A}, \mathbf{W}, \boldsymbol{\Sigma}, \mathbf{B}, \boldsymbol{\Gamma}$. Note that most parameter estimates were highly accurate, although all state covariance matrices still had to be estimated as well (i.e., with the true states provided as initialization for the E-step). Bisectrix lines in black indicate identity.

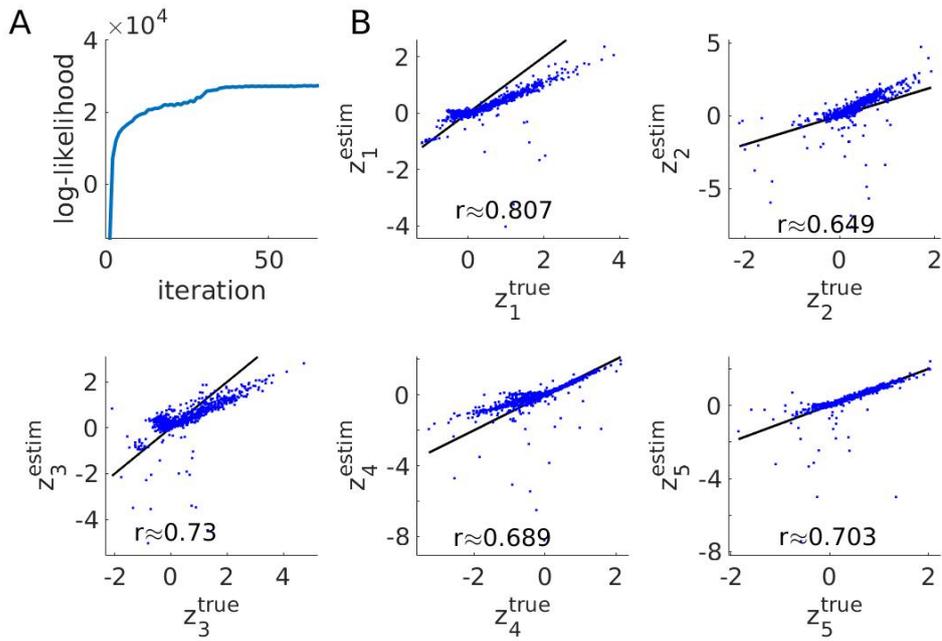

**Fig. 5.** Full EM algorithm on working memory model: State estimates for ML solution. (A) Log-likelihood as a function of EM iteration for the highest-likelihood run out of all 240 initializations. As in this example, the log-likelihood, although generally increasing, was not always monotonic (note the little ripples; see discussion in Results). (B) In this example, true and estimated states were nicely linearly related, although not with a regression slope of 1 (in general, as discussed in the text, the sets of true and estimated states may be related by some linear transformation). State estimation in this case was performed by inverting only the single constraint corresponding to the largest deviation on each iteration (see Methods). Bisectrix lines in black indicate identity.

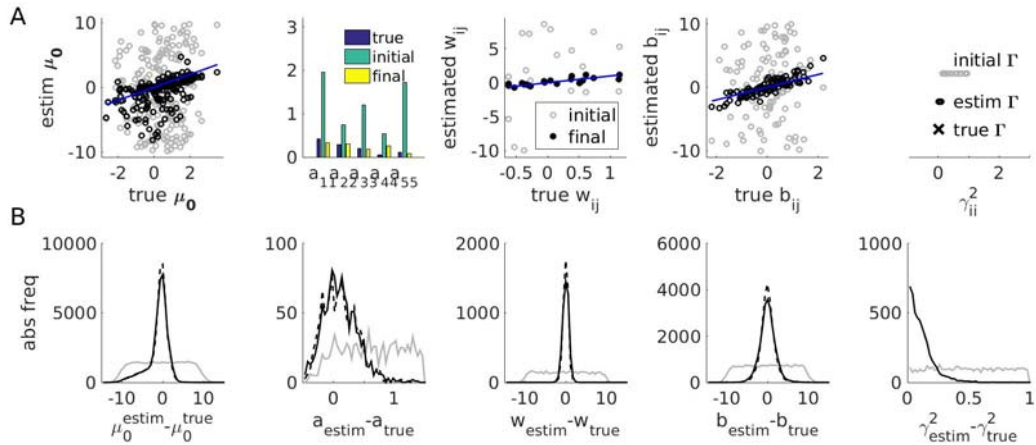

**Fig. 6.** Full EM algorithm on working memory model. (A) Parameter estimates for ML solution from Fig. 5. True parameters (on *x*-axes or as blue bars, respectively), initial (gray circles or green bars) and final (black circles or yellow bars) parameter estimates for (from left to right) $\boldsymbol{\mu}_0, \mathbf{A}, \mathbf{W}, \mathbf{B}, \boldsymbol{\Gamma}$.

Bisectrix lines in blue. (B) Distributions of initial (gray curves), final (black-solid curves), and final after reordering of states (black-dashed curves), deviations between estimated and true parameters across all 240 EM runs from different initial conditions. All final distributions were centered around 0, indicating that final parameter estimates were largely unbiased. Note that partial information about state assignments was implicitly provided to the network through the unit-specific inputs (and, more generally, may also come from the unit-specific thresholds $\theta_i$, although these were all set to 0 for the present example), and hence state reordering only produced slight improvements in the parameter estimates.

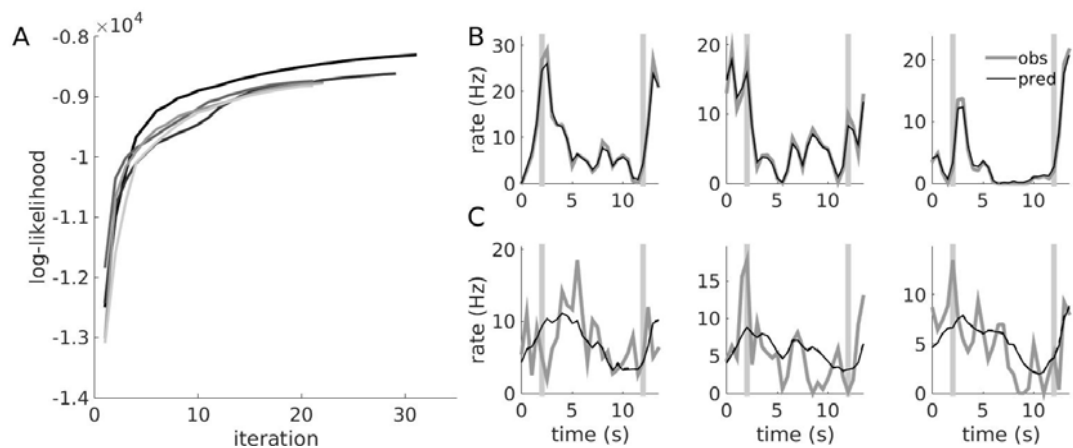

**Fig. 7.** Prediction of single unit responses. (A) Examples of log-likelihood curves across EM iterations from the 5/36 highest-likelihood runs for a 5-state PLRNN estimated from 19 simultaneously recorded prefrontal neurons on a working memory task. (B) Example of an ACC unit predicted extremely well by the estimated PLRNN despite considerable trial to trial fluctuations (3 consecutive trials shown). (C) Example of another ACC unit on the same three trials where only the average trend was captured by the PLRNN. Gray vertical bars in B and C indicate times of cue/ response. State estimation in this case was performed by inverting only the single constraint corresponding to the largest deviation on each iteration (see Methods).

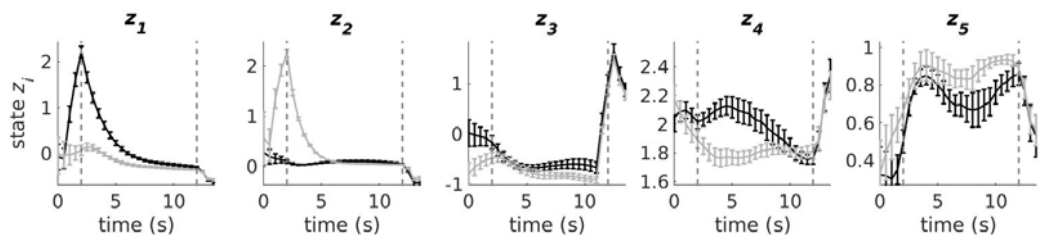

**Fig. 8.** Example for latent states of PLRNN estimated from ACC multiple single-unit recordings during working memory (cf. Fig. 7). Shown are trial averages for left-lever (black) and right-lever (gray) trials with SEMs computed across trials. Dashed vertical lines flank the 10 s period of the delay phase used for model estimation. Note that latent variables $z_4$ and $z_5$, in particular, differentiate between left and right lever responses throughout most of the delay period.

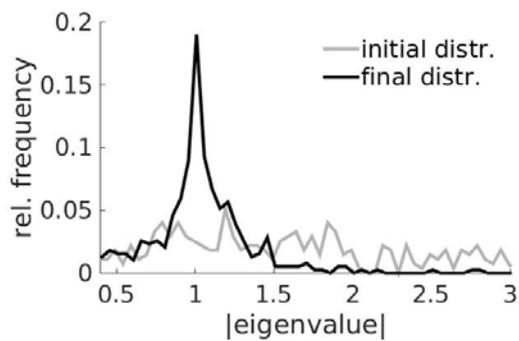

**Fig. 9.** Initial (gray) and final (black) distributions of maximum (absolute) eigenvalues associated with all fixed points of 200 PLRNNs estimated from the experimental data (cf. Figs. 7 & 8) with different initializations of parameters, including the (fixed) threshold parameters $\theta_i$. Initial parameter configurations were deliberately chosen to yield a rather uniform distribution of absolute eigenvalues $\leq 3$.